# Creating Continuously Graded Microstructures with Electric Fields via Locally Altering Grain Boundary Complexions

Qizhang Yan [1], Chongze Hu [2,3], Jian Luo [1,2,*]

[1] Department of Nano and Chemical Engineering; [2] Program in Materials Science and Engineering, University of California San Diego, La Jolla, California 92093, U.S.A.
[3] Department of Aerospace Engineering and Mechanics, University of Alabama, Tuscaloosa, Alabama 35487, U.S.A.

**Abstract**

Tailoring microstructures represents a daunting goal in materials science. Here, an innovative proposition is to utilize grain boundary (GB) complexions (*a.k.a.* interfacial phases) to manipulate microstructural evolution, which is challenging to control via only temperature and doping. Herein, we use ZnO as a model system to tailor microstructures using applied electric fields as a new knob to control GB structures locally via field-driven stoichiometry (defects) polarization. Specifically, continuously graded microstructures are created under applied electric fields. By employing aberration-corrected scanning transmission electron microscopy (AC STEM) in conjunction with density functional theory (DFT) and *ab initio* molecular dynamics (AIMD), we discover cation-deficient, oxygen-rich GBs near the anode with enhanced GB diffusivities. In addition, the field-driven redistribution of cation vacancies is deduced from a defect chemistry model, and subsequently verified by spatially resolved photoluminescence spectroscopy. This bulk stoichiometry polarization leads to preferential formation of cation-deficient (oxidized) GBs near the anode to gradually promote grain growth towards the anode. This mechanism can be utilized to create continuously graded microstructures without abnormal grain growth typically observed in prior studies. This study exemplifies a case of tailoring microstructural evolution via altering GB complexions locally with applied electric fields, and it enriches fundamental GB science.

[*] Corresponding Author. Email: jluo@alum.mit.edu (Jian Luo)

# Graphical Abstract

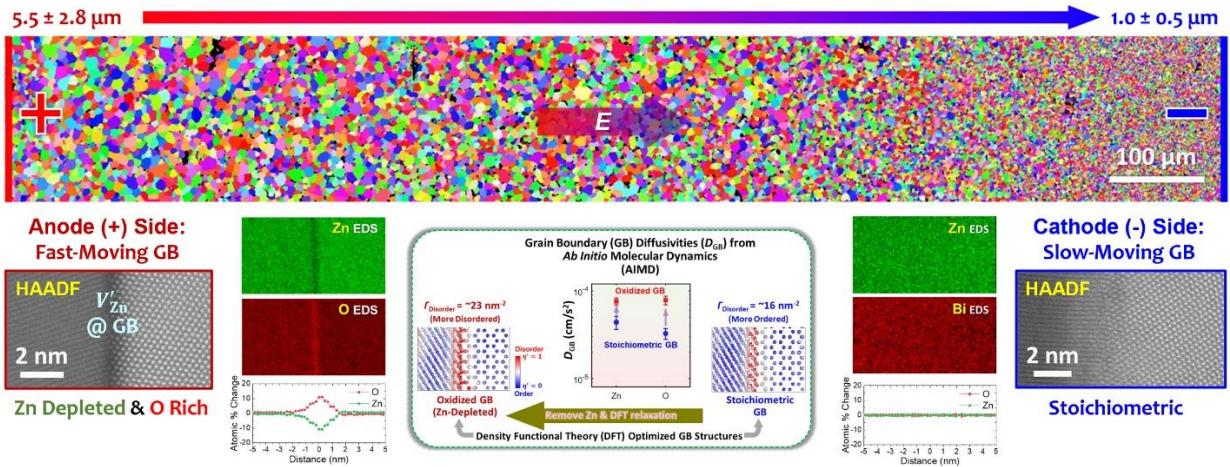

# Highlights

- Electric fields are employed to create continuously graded microstructures.
- A grain boundary oxidation mechanism for enhancing grain growth has been discovered.
- Electric fields locally alter grain boundaries through stoichiometry polarization.
- This study uncovers a new mechanism for how electric fields influence grain growth.
- It sets an exemplar of tailoring microstructures with grain boundary complexions.

## 1. Introduction

The endeavor to tailor microstructures represents a significant objective in material science. Here, an emerging concept is to utilize grain boundary (GB) phases, referred to as "complexions" to distinguish them from bulk phases [1–6], to manipulate microstructural evolution [1,4,7]. Specifically, it was hypothesized that GBs can undergo first-order or continuous transitions via changing thermodynamic potentials (*e.g.*, temperature and chemical potentials) to affect grain growth, particularly inducing abnormal grain growth (AGG) [1,4,7–9]. Yet, utilizing GB complexions to tailor microstructures meticulously by controlling only doping schemes and temperature profiles remains challenging. This motivates the current study to use applied electric fields as a new tool to locally alter GB complexions (through field-driven stoichiometry polarization) for tailoring microstructural evolution and subsequently creating *continuously graded microstructures*. In a broader perspective, this study sets an exemplar of controllably tailoring microstructures with electric fields through GB complexions.

Continuously graded microstructures have been created for metals, *e.g.*, via surface mechanical treatments [10]. On the one hand, creating such continuously graded microstructures meticulously for ceramics is more challenging. On the other hand, it has long been known that applied electric fields can influence grain growth in ceramics (with various proposed mechanisms [11–19]), thereby offering a route to create graded microstructures. It has been shown that applied direct currents can lead to asymmetric microstructures on the anode and cathode sides, but often result in undesirable abrupt or abnormal grain growth [13,15,19,20]. Using blocking electrodes, Rheinheimer *et al.* also showed increased GB mobility at the single crystal-polycrystal interface near the cathode in $SrTiO_3$ in the absence of any current (explained by the redistribution of oxygen vacancies and associated solute-drag effects, with no observable GB structural transition) [14,21,22]. Yet, no prior study was dedicated to *intentionally* utilizing electric fields or currents to tailor microstructures in a controlled manner, in particular, to create continuously graded microstructures with electric fields via locally altering GB complexion.

Notably, a recent study discovered that an applied electric field could alter grain growth by inducing a GB transition [19]. In this case, the reduction of redox-active Bi cations segregated at GBs in $Bi_2O_3$-doped ZnO was found to cause a unique GB disorder-to-order transition that can subsequently induce AGG near the cathode [19,23]. This GB transition stems from the $Bi_2O_3$-segregation-induced formation of amorphous-like intergranular films (disordered GBs) and subsequent interfacial ordering upon (cathode-side) Bi reduction [19], which is a material-specific mechanism. Unfortunately, the enhanced grain growth is discontinuous (resulting in AGG) in this case, and continuously graded microstructures cannot be achieved, presumably because of the strong redox activity of segregated Bi cations. Although this study [19] pointed to the possibility of altering microstructural evolution through inducing GB transitions with applied electric fields, more general and gentler GB transition mechanisms are needed to enable the controlled generation of continuously graded microstructures.

Herein, we use undoped ZnO as a model system to show the feasibility of creating continuously graded microstructures with electric fields via altering GB complexions locally via field-driven stoichiometry (defects) polarization in a controlled manner. By coupling aberration-corrected scanning transmission electron microscopy (AC STEM) with density functional theory (DFT) and *ab initio* molecular dynamics (AIMD), we reveal an anode-side GB oxidation transition to increase GB diffusivity. We further show that field-driven redistribution of cation vacancies results in preferential GB oxidation to gradually enhance grain growth towards the anode. This newly discovered anode-side GB oxidation transition, which arises from GB enrichment of cation vacancies, is likely more general and more controllable. This discovery not only enables the creation of continuously graded microstructures with applied electric fields via altering GB complexions locally through field-driven stoichiometry polarization, but also advances the emerging GB complexion theory [24–28] to enrich the fundamental GB science.

## 2. Results and discussion

### 2.1 Graded microstructures

Using a simple undoped ZnO as our model system, we first demonstrated the creation of continuously graded microstructures via an applied electric field. An example of a continuously graded microstructure created via annealing a ZnO polycrystalline specimen under an applied electric field is shown in Figure 1. In this case, a constant voltage of 3.84 V was applied onto a dense polycrystal ZnO specimen of ~1.43 mm in thickness (producing a nominal electric field of ~26.9 V/cm and an electric current density of 295 mA/mm$^2$) under a furnace temperature of 600 °C for 4 h. Figure 1e visually shows the continuous grain size variation in the graded microstructure. The ZnO grain sizes were found to gradually increase from the negative electrode (cathode) to the positive electrode (anode). Electron backscatter diffraction (EBSD) mapping was carried out to quantify the grain size variation across the specimen. The average grain size (equivalent grain diameter) near the anode was measured to be 5.5 ± 3.8 μm, which is about 5 times larger than the grain size of 1.0 ± 0.5 μm near the cathode. The large grain size gradient was unlikely caused by any temperature gradient, given a thin pellet (<1.5 mm). As shown in Supplementary Figure S2, anode-side discontinuous grain growth was observed only in one experiment with the lowest specimen temperature of ~1060 °C; graded microstructures have been achieved in all other cases with estimated specimen temperatures in the range of ~1180°C to ~1480 °C. Supplementary Figures S3-S8 document six graded microstructures created under different experimental conditions.

The graded microstructure can be controlled through changing the applied electric field. As shown in Figure 1f, a smaller grain size variation was observed at a lower nominal applied electric field (*E*) of ~20.2 V/cm (*vs. E* = ~26.9 V/cm for the baseline case shown in Figure 1e) at the same furnace temperature of 600 °C. The measured grain size *vs.* normalized location curves for these two cases are compared in Figure 1h. However, we should note that the sample temperatures in these two cases were different due to Joule heating. They were estimated to be ~1380 °C (for *E* =

~20.2 V/cm) and ~1214 °C (for $E$ = ~26.9 V/cm), respectively. To exclude the possible temperature effect, Figure 1g further illustrates an example of a true field effect, where the sample temperatures were kept similar (~1400 °C) via carefully tuning the furnace temperature to compensate the different Joule heating. Comparing Figure 1e and Figure1g, a higher grain size gradient and a large anode-side grain size were evident for the higher applied electric field (~26.9 V/cm *vs.* ~24.3 V/cm), even at a slightly lower estimated specimen temperature (~1380 °C *vs.* ~1408 °C). Thus, this comparison (Figure 1i) shows a true field effect.

## 2.2 Grain growth kinetics under the field

To quantify the grain growth kinetics, the microstructure evolution of ZnO specimens annealed under a constant applied voltage of 2.80 V for different durations was studied (Figure 2). The pristine dense ZnO specimen fabricated by spark plasma sintering (SPS) showed a slightly larger grain sizes at the center of the specimen compared to the surface regions. Increasing the annealing time under the same electric field gradually increased grain sizes from the anode side, with little changes in grain sizes at the cathode side, as shown in Figure 2d. For the sample annealed for 32 hours (Figure 2a), the anode-side grain size increased to $4.8 \pm 2.3$ μm, which is significantly larger than the cathode-side grain size ($1.9 \pm 0.9$ μm).

We can fit the local grain size kinetics to a power law:

$$d_t^n - d_0^n = kt, \tag{1}$$

where $d_t$ and $d_0$ are the grain sizes at time $t$ and $t = 0$, respectively, $n$ is the grain growth exponent, and $k$ is a kinetic constant. For undoped ZnO, $n = 3$ was consistently observed in the previous grain growth studies [29,30]. Using the anode- and cathode-side grain sizes *vs.* time curves shown in Figure 2e, the anode-side kinetic constant was fitted to be $k_{anode}$ = ~3.8, which is more than 30 times greater than that at the cathode side ($k_{cathode}$ = ~0.11). Thus, we can conclude that the GB mobility, which should be scaled by $k$, was greatly enhanced at the anode side.

## 2.3 GB oxidation transition

To elucidate the atomic-level mechanism of the anode-side GB mobility enhancement, the GB structures were examined with AC STEM, in conjunction with energy dispersive X-ray spectroscopy (EDS). Nine (9) randomly selected GBs were lifted out using a focus ion beam (FIB) from the cathode and the anode sides of the ZnO specimen after annealing under an applied electric field for AC STEM and EDS characterization.

Five (5) GBs randomly selected from the anode side (Figure 3a-3d, Supplementary Figure S9) all show similar characteristic dark contrasts in the AC STEM high-angle annular dark-field (HAADF) images, which suggests Zn deficiency (as Zn is the heavier element with a higher Z number in comparison with O). STEM EDS elemental mapping was conducted to further examine the GB composition. Figures 3e and 3f show the Zn and O maps of an anode-side GB, which directly show that the GB is Zn-poor and O-rich. Supplementary Figure S9 further documented the STEM HAADF image and EDS Zn and O maps for another (the fifth randomly selected) GB at the anode size, which also showed similar characteristic dark contrast in HAADF image. The STEM ESD elemental maps again directly confirmed that this GB is also Zn-deficient and O-rich. The assembly of the characterization results from five randomly selected GBs showed a consistent result of the formation of Zn-deficient (O-rich) GBs at the anode side with characteristic dark contrasts in HAADF images.

In contrast, stoichiometric GBs were observed for all four (4) GBs randomly selected from the cathode side, as shown in the HAADF images in Figure 3h-3j. The characteristic dark contrast of the Zn-deficient GBs was not observed in any of the four randomly selected GBs from the cathode side. EDS Zn and O maps (Figure 3l and 3m) further confirmed that the GB is stoichiometric.

Furthermore, the compositional line profile was extracted across the GB, which shows O enrichment and Zn depletion at the anode-side GB quantitatively (Figure 3g). For the anode-side GBs, the O composition increased by ~11% at the GB region compared to the bulk region, and the Zn composition reduced by the same amount. Given the possible smearing effect of the electron beam in STEM, the actual GB nonstoichiometry (O:Zn ratio) can be even higher than that shown by the line profile in Figure 3g. In contrast, there was no obvious change in O and Zn composition across the GB at the cathode side (Figure 3n).

We should emphasize that all nine (9) GBs were randomly selected from the polycrystalline specimen (prepared using a focused ion beam). We have HAADF images for five (5) randomly selected GBs (Fig. 1a-d and Suppl. Fig. S9) on the anode side, all of which exhibit the same characteristic dark contrasts associated with Zn deficiency at the GBs. ESD elemental mapping directly confirmed that two of them are Zn-deficient and O-rich (Fig. 1e-f and Suppl. Fig. S9). In strong contrast, HAADF images for four (4) randomly selected GBs on the cathode side lack such characteristic dark contrasts at the GBs (associated with Zn deficiency). ESD elemental mapping also directly confirmed that these GBs are stoichiometric (neither Zn-deficient nor O-rich). The compilation of these results from nine (9) randomly selected general GBs provides a statistically significant conclusion: the formation of Zn-deficient, O-rich GBs on the anode side versus stoichiometric GBs on the cathode side.

In summary, the microstructure and GB characterization results collectively suggest that oxidized GBs (with Zn deficiency) formed near the anode, which led to enhanced grain growth. This resulted from preferential GB oxidation near the anode due to the field-driven stoichiometry polarization that altered GB structures locally. Further analyses of defects and a GB model, as well as validation by spatially resolved photoluminescence spectroscopy, suggest that the field-driven defects redistribution (enrichment of Zn vacancies near the anode) can cause preferential Zn vacancy segregation to form such oxidized GBs at the anode side.

## 2.4 DFT and AIMD simulations

To verify the enhancement of GB kinetics in oxidized GBs and further investigate the underlying mechanism, we perform first-principles DFT calculations and subsequently AIMD simulations based on the DFT-relaxed GB structures.

Based on the STEM images, we constructed three asymmetric GB structures to represent general GBs for DFT calculations. For each model GB, we typically set one grain terminal plane to mimic a low-index grain surface observed in a STEM image and selected the other matching grain surface to satisfy a periodic boundary condition to enable DFT and AIMD calculations. We selected asymmetric GBs to represent more general GBs. The details of the three model GBs selected are documented in the Materials and Methods section. For each model GB, we first obtained a DFT-relaxed interfacial structure for the stoichiometric GB. Subsequently, we removed Zn atoms from the GB region and relaxed the interfacial structure by DFT again to represent the Zn-deficient, oxidized GB for each case. Note that we could alternatively obtain similar results for the oxidized GBs via adding O atoms and conducting DFT relaxation, which produced consistent results (Supplementary Figure S10).

The DFT-optimized stoichiometric and oxidized (*a.k.a.* Zn-deficient) GB structures of the first asymmetric GB (denoted as "GB I") are shown in Figure 4a and 4b, respectively. The corresponding maps of a disorder parameter (based on an order parameter proposed by Chua *et al.* [19,31]) are illustrated in Figure 4c and 4d, respectively. Subsequently, GB excess of disorder ($\Gamma_{\text{Disorder}}$) and interfacial width were calculated following the methods used in Ref. [19]. Notably, the Zn-deficient, oxidized GB exhibits a more disordered structure with a larger $\Gamma_{\text{Disorder}}$ of ~ 23.2 nm$^{-2}$ and a larger interfacial width of ~0.79 nm (Figure 4d). The stoichiometric GB has a less disordered structure with a relatively smaller $\Gamma_{\text{Disorder}}$ of ~16.5 nm$^{-2}$ and a smaller interfacial width of ~0.61 nm (Figure 4c). This observation agrees with STEM images shown in Figure 3, where oxidized GBs near the anode side appear to be wider in the effective interfacial width and more disordered than those stoichiometric GBs near the cathode. In addition, the differential charge density maps were calculated (Figure 4e and 4f), which again indicate that the stoichiometric GB has a more ordered charge transfer pattern. To show the generality of the results, two additional asymmetric GBs (GB II and GB III) were relaxed by DFT (Figure 4h and Supplementary Figure S10).

The large-scale AIMD or *ab initio* molecular dynamics simulations were subsequently applied to calculate and compare the GB diffusivities for DFT-relaxed stoichiometric *vs.* oxidized GBs for all three asymmetric GBs (>200 atoms in each simulation cell). Figures 4g shows the calculated GB diffusivities of Zn and O atoms in both stoichiometric and oxidized GBs at 1573 K for GB I. To show the generality, similar AIMD simulations were performed for GB II and GB II, and simulated GB diffusivities are shown in Figure 4h and 4i.

Taking GB I as an example, we plotted mean squared displacement (MSD) and potential energy vs. simulation time curves in Supplementary Figures S11 and S12, respectively. As shown in Supplementary Figure S12, the potential energy of GB I converged and reached a plateau, indicating that the GB structure reached equilibrium. The trends in the MSD plots (Supplementary Figure S11) also show a gradual increase as a function of time, further supporting that the GB approached equilibrium. The MSD and potential energy *vs.* simulation time plots for GB II and GB III (Supplementary Figures S13-S16) exhibit similar trends. For each AIMD simulation, we calculated the average GB diffusivities of Zn and O atoms based on five different trajectory paths for each GB structure. We then replotted the standard deviations as error bars in Figure 5. The small standard deviations from the five different trajectory paths also suggest that the AIMD simulation results are highly repeatable.

In all three cases, the GB diffusivities of both Zn and O atoms are markedly higher in the Zn-deficient, oxidized GBs than those in stoichiometric GBs (Figure 4g-4i). The consistent simulation results from all three GBs suggest that the oxidized (Zn-deficient) GBs have high GB diffusivities, which agrees well with the experimental observations that oxidized GBs at the anode side exhibited higher mobility (that should scale with GB diffusivities) that led to enhanced grain growth.

## 2.5 Stoichiometry (defects) polarization

It is well known that applied electric fields can cause stoichiometry (defects) polarization in ionic oxides, leading to gradients in the concentrations and chemical potentials of charged defects (*e.g.*, oxygen vacancies) due to electrochemical coupling effects (see, *e.g.*, Waser, Baiatu, and Härdtl's classical work on $SrTiO_3$ [32,33]). Such defect gradients are known to result in abrupt grain growth on the cathode side in $Y_2O_3$-stabilized $ZrO_2$ (YSZ) with large current densities [13,15] and enhanced GB mobility at the single crystal-polycrystal interface in $SrTiO_3$ with blocked electrodes (negligible currents) [14]. To analyze the field-driven redistribution of charged defects in ZnO in our case, we adopted a defect chemistry model built by Sukkar and Tuller [34–36]. In undoped ZnO, the Frenkel defects dominate over the Schottky defects so that zinc interstitials are the dominant donor defects in air (at $P_{O2} = 0.2$ atm) [34–36]. Zinc vacancies can dominate at lower $P_{O2}$ region (*i.e.*, near the anode due to the electrochemical reduction in our specimens), as well as form preferentially at GBs [34,37].

Figure 5a shows a Brouwer diagram from the Sukkar-Tuller model [36], where the concentrations of defects are plotted as a function of oxygen partial pressure ($P_{O2}$). At high $P_{O2}$ (*i.e.*, in the oxidized region near the anode in our specimens), ZnO is a *p*-type semiconductor with the charge neutrality condition: $p = [V_{Zn}']$ (in the Kröger-Vink notation). At low $P_{O2}$, ZnO is a *n*-type semiconductor where negatively charged electrons are balanced by Zn interstitials with double and single positive charges in two successive regions towards the cathode (with the charge neutrality conditions of $n = 2[Zn_i^{\bullet\bullet}]$ and $n = [Zn_i^{\bullet}]$, respectively). Between the *p* and *n* regions, an ionic region can exist within a narrow $P_{O2}$ range with the charge neutrality condition of $[V_{Zn}'] = 2[Zn_i^{\bullet\bullet}]$. However, conduction may still occur though electrons and holes in our specimens under electric fields between the *p*- and *n*-regions to maintain the continuity (that will be discussed subsequently) in this nominally "ionic (*i*) region".

We can estimate the electrostatic potential *vs.* $P_{O2}$ relation in the *p* and *n* regions based on the Brouwer diagram shown in Figure 5a. For a given charged specie *i*, the flux ($J_i$) and the corresponding current density ($j_i$) are proportional to the driving force:

$$j_i = zFJ_i = -\frac{\sigma_i}{zF}\frac{d\eta_i}{dx} = -\frac{\sigma_i}{zF}\left(\frac{d\mu_i}{dx} + zF\frac{d\phi}{dx}\right) \tag{2}$$

where $x$ is the spatial variable, $\sigma_i$ is the conductivity, $z$ is the charge of the mobile specie, $F$ is the Faraday constant, $\eta_i$ ($\equiv \mu_i + zF\phi$) is the electrochemical potential, $\mu_i$ is the chemical potential, and $\phi$ is the electrostatic potential.

ZnO is known as an electronic conductor. In addition, the ZnO specimen was placed between two ion-blocking Pt electrodes during our grain growth experiment under an applied electric field. Thus, we assume no ionic current at the steady state. For an ionic specie, $J_i = 0$ at the steady state gives:

$$\frac{d\mu_i}{dx} = -zF\frac{d\phi}{dx}. \tag{3}$$

Integration of Eq. (3) produces:

$$\Delta\phi = -\frac{1}{ZF}\Delta\mu_i = -\frac{RT}{zF}\ln\left[\frac{C_i(x_2)}{C_i(x_1)}\right] \tag{4}$$

where $R$ is the gas constant, $T$ is specimen temperature, and $C_i(x)$ is the defect concentration at location $x$. Here, we assume that defects form ideal (dilute) solutions. Then, we estimated the electrostatic potential drops ($\Delta\phi$) for different segments shown in Figure 5a based on the ratios of dominating ionic defects concentrations [$V_{Zn}'$], [$Zn_i^{\bullet\bullet}$], and [$Zn_i^{\bullet}$], respectively. We labeled the calculated $\Delta\phi$ values on the top of Figure 5b. Comparing with the $P_{O2}$ values on the bottom of Figure 5a, we could then plot the estimated $P_{O2}$ vs. the electrostatic potential curve across the specimen in Figure 5b. Here, we estimated the $\Delta\phi$ for the ionic region based on an equilibrium of the (presumably more mobile) $V_{Zn}'$ vacancies (while noting the conduction at the *p-i-n* junction can be more complicated and will be discussed later).

Figure 5b suggests that applied voltages of ~2.8-3.8 V (the typical range that led to the creation of excellent graded microstructures in our study) can lead to extreme redox conditions ($P_{O2} \sim 10^{\pm 20}$ to $10^{\pm 30}$ atm) at the anode and the cathode (albeit we do not know the exact $P_{O2}$ values at the two specimen surfaces due to the unknown boundary conditions, as elaborated below). The occurrence of ionic conduction (if any) can reduce the $P_{O2}$ range. Yet, Figure 5b clearly shows the extreme local oxygen activities that can be achieved by applying a couple of volts on the ZnO specimens. In the current case, it led to anode-side GB oxidation to create graded microstructures.

The available data (*e.g.*, the Sukkar-Tuller model [36] that was developed based on limited experimental data that are still under scrutiny and debates [38–43],) is not sufficient for accurately solving the defects distributions in ZnO, which will also be affected by the more conductive GBs [36,37] that are not considered in the bulk defects models. Moreover, if both electrodes were ion-blocking, neither of them was in equilibrium with the atmosphere ($P_{O2} = 0.2$ atm). Consequently, we cannot obtain the exact $P_{O2}$ values at the two surfaces of the specimen. Thus, we could not provide the actual $p_{O2}$ profile *vs.* location from anode to cathode. Instead, we plotted the estimated oxygen partial pressure in the specimen as a function of electrostatic potential. We put $p_{O2} = 0.2$ atm at the middle of graph (which may not correspond to the middle of the specimen). The actual $p_{O2}$ vs. distance profile may shift. Nonetheless, some analyses can be made to identify useful and valid trends for discussion.

At the steady state, a forward biased *p-i-n* junction may form (Figure 5b) for the bulk ZnO structure. For simplicity, we can assume that the steady-state current density ($j_{SS}$) is constant throughout the specimen (with no ionic current in *p* and *n* regions):

$$j_{SS} = j_h \text{ (in the } p \text{ region)} = j_e \text{ (in the two } n \text{ regions)} \tag{5}$$

In the *p* region near the anode, the dominate defects are $V'_{Zn}$ and holes. We assume that both the charge neutrality condition ($p = [V'_{Zn}]$ so that $d\mu_{V'_{Zn}} = d\mu_h$) and defect equilibria hold locally. For $V'_{Zn}$ ions ($z = -1$), we can derive from Eq. (3):

$$F\frac{d\phi}{dx} = \frac{d\mu_{V'_{Zn}}}{dx} = \frac{d\mu_h}{dx}. \tag{6}$$

The steady current density is given by:

$$j_{SS} = j_h = -\frac{\sigma_h}{F}\left(\frac{d\mu_h}{dx} + F\frac{d\phi}{dx}\right) = -2\sigma_h\frac{d\phi}{dx}, \tag{7}$$

which suggests the electric field follows $E = -d\phi/dx = j_{SS}/(2\sigma_h)$ in the *p* region. Likewise, we can derive $E = j_{SS}/(5\sigma_e)$ in the intermediate *n* region and $E = j_{SS}/(2\sigma_e)$ in the *n* region near the cathode following similar assumptions. Note that we cannot determine the exact location of the *p-i-n* junction due to the unknown boundary conditions ($P_{O2}$ values) at the two surfaces of the specimen.

The conduction in the nominally ionic (*i*) region can be more complicated. Conduction through the *p-i-n* junction may still occur through electrons and holes (as they drift into the *p-i-n* junction and recombine in the nominally ionic region). This is because (1) electron and hole concentrations are only slightly lower than those of ionic species and (2) converting an electronic current to an ionic current would need internal defect reactions to generate and annihilate the ionic species to ensure mass conservation. Nonetheless, Eq. (3) cannot be simultaneously satisfied for both $V'_{Zn}$ vacancies and $Zn_i^{\bullet\bullet}$ interstitials while maintaining local charge neutrality ($[V'_{Zn}] = 2[Zn_i^{\bullet\bullet}]$). This can be circumvented by breaking the local charge neutrality (to form a space charge zone at the *p-i-n* junction) and/or by deviating from the local defects equilibria. It is also possible to generate an ionic current with internal defects reactions to obey both mass and charge balances in conservative ensembles [44–47]. Similar *p-i-n* junctions have been derived and modeled by Dong and Chen [15,48–51] and Kirchheim [52] for ion-conducting YSZ and $CeO_2$ (with different dominant defects).

Since the hole and electron conductivities ($\sigma_h$ and $\sigma_e$) depend on the carrier concentrations (*p* and *n*), the electric field is not a constant (so that the applied electric fields in our experiments are nominal or represent average values across the specimens). The steepest electrostatic potential drop (*i.e.*, the highest local field) is likely at the *p-i-n* junction where the carrier concentrations (*p* and *n*) are the lowest (Figure 5a). The existence of similar oxygen (and electrostatic) potential transitions was previously revealed for YSZ, which, however, resulted in abrupt (abnormal) grain growth at the cathode side [13,15,48–51] that differs from the current case. Notably, the current case showed gradual increases in the grian sizes towards the anode side without abrupt or abnormal grain growth (with only one exception shown in Supplementary Figure S2 that will be discussed and explained later). Several factors can contribute to the desirable continuous variations in grain sizes in undoped ZnO. First, a prior numerical analysis showed that adding an extra conductivity can reduce the oxygen potential gradients at the *p-n* junction [51]. For ZnO, GBs are more conductive (as suggested by the size dependent conductivity [53] and analogous the well-known surface excess of carriers [39]), which can contribute as an extra conductivity that reduce the gradients at the bulk *p-i-n* junctions for polycrystalline ZnO specimens. Second, high specimen temperatures can substantially reduce the minimum $\sigma_h$ and $\sigma_e$ to reduce the gradients at *p-i-n* junctions, which is consistent with our experimental observation that abrupt grian growth only took place in the case of the lowest specimen temperature (Supplementary Figure S2). Third, the existence of two *n* regions, where the electrical field will reduce by 2.5× (from $j_{SS}/(2\sigma_e)$ to $j_{SS}/(5\sigma_e)$) moving towards the anode, will also reduce the average gradient in the *n* regions.

It is also noted that the cathode-side AGG in YSZ and $CeO_2$ occurs through a bulk mechanism (reduction-enhanced bulk diffusion) [15,48–51]. In the current case of undoped ZnO, we propose a different cation-vacancy-induced GB oxidation mechanism that can be more continuous, which will be discussed subsequently.

**2.6 The GB model**

Tuller also proposed the GBs in ZnO can preferentially oxidize to form $V'_{Zn}$ (and even $V''_{Zn}$) when the zinc vacancies are not the dominant defects in air [34,36]. We can explain this with a simple phenomenological model assuming preferential GB segregation of $V'_{Zn}$ vacancies:

$$\frac{\Gamma^{GB}_{V'_{Zn}}}{\Gamma_0 - \Gamma^{GB}_{V'_{Zn}}} = \frac{[V'_{Zn}]}{1-[V'_{Zn}]} \cdot e^{-\frac{\Delta g^{seg.}_{V'_{Zn}}}{kT}}, \tag{8}$$

where $\Gamma^{GB}_{V'_{Zn}}$ is the amount of "$V'_{Zn}$ vacancies" at the GB core (that is put in quotation marks because they should possess a different character than the well-defined vacancies in the bulk ZnO structure due to a different bonding environment at GBs), $\Gamma_0$ is the maximum amount, and $\Delta g^{seg.}_{V'_{Zn}}$ is the free energy of segregation per $V'_{Zn}$ vacancy. Without applying an electric field, the $[V'_{Zn}]$ in the bulk phase is low (in air, as shown in Figure 5a), but $\Gamma^{GB}_{V'_{Zn}}$ can be appreciable according to Tuller [34,36]. Thus, the effective $\Delta g^{seg.}_{V'_{Zn}}$ value is negative and significant. When an electric field is applied, $[V'_{Zn}]$ in the bulk phase can increase substantially in the anode side due to the stoichiometry polarization, which can subsequently raise $\Gamma^{GB}_{V'_{Zn}}$ at the GB core to form Zn-deficient, oxidized GBs. In the real GB structure, an individual cation vacancy may not be well-defined at the GB core (due to a different bonding environment), and its effective charge can also change (*e.g.*, by further oxidation), particularly in the presence of a high effective concentration and a more oxidized environment. Instead, the GB core should undergo a structural transition to become Zn-deficient and O-rich, which is also likely to become more disordered. The realistic GB structures can be better modeled by the DFT relaxation (Figure 4). In general, we expect more disordered GBs to have high kinetic rates [1,7,54–57] and non-stoichiometry (with cation-vacancy-like structures at the GB core) can also help diffusion. This was indeed verified by our AIMD simulations (Figure 4). We also expect that (1) cation-vacancy-induced GB oxidation can occur more continuously and gently compared to the redox of segregants (*e.g.*, the reduction of segregated $Bi^{3+}$ at ZnO GBs [19]), and (2) gradual GB oxidation can take place before electrochemically induced bulk oxidation. Both of these factors can contribute to smoother transitions and subsequently more gradual variations in grain sizes.

In addition, the preferential GB oxidation can also cause a negatively charged GB core, which will lead to the enrichment of holes in the space charge region near the GB, as schematically illustrated in Figure 5c. This can also result in increased GB conductivities (that can also reduce the potential gradients at the *p-i-n* junction, as we discussed earlier).

## 2.7 Validation by photoluminescence spectroscopy

To probe the defect concentrations across the specimen to validate the above prediction from the defect chemistry model, we used spatially resolved photoluminescence spectroscopy to quantify Zn vacancies (the dominant defects at the anode side, predicted by the model) across the ZnO specimen annealed under a constant voltage of 3.8 V for 4 h, as shown in Figure 5d-g. It is known that Zn vacancies in ZnO give rise to the green luminescence centered around 2.35 eV or a wavelength of 530 nm (albeit the valence state of Zn vacancies cannot be resolved) [58–60]. The map of the photoluminescence intensity at the 530 nm wavelength shown in Figure 5d clearly indicates higher concentrations of Zn vacancies towards the anode. We can also visualize this trend in the averaged photoluminescence spectra as a function of normalized distance to the anode (Figure 5e). For example, the peak intensity of the green emission (from Zn vacancies) near the anode side is more than twice higher than that near the cathode side (Figure 5g). This observation fully agrees with the defect polarization prediction from the defect chemistry model. The increased green emission near the anode may also include the contribution from GBs (enriched in Zn vacancies), in addition to that from the Zn vacancies in the bulk phase.

Differing from YSZ, $CeO_2$, and $SrTiO_3$, oxygen vacancies are not the dominant defects in any region in ZnO in the current case. However, the $P_{O2}$ (or oxygen vacancy concentration) changes drastically from anode to cathode, depending on the concentration of dominant Zn defects ($[V_{Zn}']$, $[Zn_i^{\bullet\bullet}]$, and $[Zn_i^{\bullet}]$). Thus, we need to analyze and probe the concentration of dominant Zn defects (*e.g.*, $[V_{Zn}']$ near the anode) to determine the local $P_{O2}$ (redox environments, as the concentration of oxygen vacancies in the bulk structure is too low to be directly detectable.

## 2.8 General Discussion

While it is long known that applied electric fields can alter grain growth in ceramics [11–19], the current study represents the first effort to *intentionally* use applied electric fields to create continuously graded microstructures, which is realized via oxidizing GBs locally as a result of field-driven stoichiometry polarization, a new mechanism discovered in this study. This study also represents one of very few cases where the underlying mechanisms of the field effects on grain growth have been revealed, and the mechanism uncovered here differs from those reported previously [13–15,19]. Notably, this cation-vacancy induced GB oxidation mechanism discovered here in undoped ZnO differs from the Bi-redox induced GB disorder-to-order transition in $Bi_2O_3$-doped ZnO [19] and bulk reduction enhanced grain growth in YSZ and $CeO_2$ that is not due a GB transition [13,15,50], both of which resulted in undesirable abrupt and abnormal grain growth at the cathode side. Moreover, this cation-vacancy induced GB oxidation mechanism is likely more general and controllable. We also note that prior studies on $SrTiO_3$ [14,61] focusing on weak fields and used current-blocking electrodes (with negligible current), which produced different field effects at different conditions from the current experiments. Rheinheimer and Hoffmann observed increased GB mobility at the single crystal – polycrystal interface near the cathode in a well-designed experiment and explained it from the reduced solute drag of oxygen vacancies [14,21,22,61], where TEM study did not reveal any detectable change in the GB structure. Electrochemically driven AGG in $SrTiO_3$ was also studied via phase-field modeling [62]. The current study markedly differs from all these prior studies in both its goal of controllably creating continuously graded microstructures via GB transitions and a different underlying mechanism uncovered.

This study not only points to a new opportunity to tailor microstructural evolution with electric fields, but also advances the emerging GB complexion theory [1,4]. Most significantly, this work exemplifies a rare case of using GB transitions to tailor microstructures, which has been argued to represent a potentially transformative direction in microstructural engineering [1,4,9], but with very few successful examples of meticulous microstructure control (through manipulating temperature profiles and/or doping only) to date.

In a broader perspective, electric fields are widely used for innovative materials processing technologies (*e.g.*, spark plasma sintering [16,63,64] or flash sintering [53,64,65]) and energy storage and conversion devices (*e.g.*, fuel cells or solid-state batteries) [51,66,67], where they can often lead to unexpected (usually undesirable) microstructure changes and instability. Thus, the discovery of a new, general mechanism of GB oxidation to alter GB kinetics also provides a new clue to understanding some of the unresolved electric field effects on microstructural evolution and instability that have puzzled materials scientists for decades.

## 3. Conclusions

In this study, we have used ZnO as a model system to show the controlled generation of continuously graded microstructures with electric fields. Moreover, we have revealed a new cation-vacancy induced GB oxidation mechanism that can alter grain growth gradually and locally through field-driven stoichiometry polarization. This work has also advanced fundamental GB science, in particular, enriched the emergent GB complexion theory. It exemplifies a case of using GB complexions to tailor microstructures, specifically creating continuously graded microstructures under applied electric fields by locally altering GB complexions through field-driven stoichiometry (defects) polarization.

## 4. Materials and methods

### 4.1 Materials preparation

ZnO powders (99.98% purity, 18 nm, US Nanomaterials) was first sintered into dense pellet using spark plasma sintering (SPS) at 840 °C for 5 min under a uniaxial pressure of 50 MPa in vacuum ($10^{-2}$ Torr) using a graphite die. The pellet was then annealed in air at 700 °C for 9 hours to remove residual carbon. All sintered pellets reached >99% of the theoretical densities. The sintered pellets were cut into pieces of the size: ~4 × ~3 × ~1.5 mm$^3$. Pt was sputtered onto both sides of the specimens to form electrodes.

### 4.2 Electric field effects on grain growth experiments

For the grain growth experiment under an applied electric field, an external direct current was applied to the specimen while maintaining a constant voltage. The grian growth experiments were conducted in a tube furnace isothermally with a home-made fixture. Tektronix DMM 4040 digital multimeters were used to record the current and electrostatic potential readings. The sample temperature was higher than the furnace temperature because of the Joule heating. The sample temperatures were estimated based on the black body radiation model [68,69].

### 4.3 Materials characterization

The cross-sectional microstructures were characterized by scanning electron microscopy (SEM) using a FEI Apreo microscope equipped with an Oxford System electron backscatter diffraction (EBSD) detector. Transmission electron microscopy (TEM) lamellas were fabricated using a Thermo Scientific Scios dual beam focused ion beam (FIB)/SEM system. The atomic level GB structures were characterized with an aberration corrected scanning TEM (AC STEM) using a JEOL JEM-300CF microscope operating at 300 kV, in conjunction with X-ray energy dispersive spectroscopy (EDS).

To study the defects polarization of the specimens annealed under applied electric fields, spatially resolved photoluminescence spectroscopy was conducted on the sample cross sections on a Leica SP5 confocal microscope. The wavelength range was from 400 to 700 nm.

### 4.4 Density functional theory (DFT) calculations

We used the GBGenerator code in Python Materials Genomics (pymatgen) library to construct ZnO GB structures for first-principles DFT calculations [70,71]. The lattice parameters of the ZnO hexagonal were set to $a$ = 3.29 Å and $c$ = 5.31 Å based on the data from the Materials Project [72].

Three types of asymmetric Zn GB structures are considered. For the first model GB (denoted as "GB I"), we fixed the lower index plane as $(11\bar{2}0)$ plane and rotated the other plane along the [100] axis with a rotation angle ~53°. This process generated an asymmetric GB with $\Sigma = 5$, and its triclinic cell has following lattice parameters: $a = 6.24$ Å, $b = 14.49$ Å, $c = 39.47$ Å, $\alpha = 109.90°$, $\beta = 74.73°$, and $\gamma = 74.73°$. The total number of atoms in simulation cell for GB I is 238.

For the second model GB (denoted as "GB II"), the lower index plane was also fixed to $(11\bar{2}0)$ plane and we rotated the other plane along the $[10\bar{1}0]$ axis with a rotation angle ~136°. This process generated an asymmetric GB with $\Sigma = 7$, and its triclinic cell has following lattice parameters: $a = 11.20$ Å, $b = 17.26$ Å, $c = 62.29$ Å, $\alpha = 29.31°$, $\beta = 151.76°$, and $\gamma = 140.72°$. The total number of atoms in the simulation cell for GB II is 221.

For the third model GB (denoted as "GB III"), the lower index plane was fixed to $(01\bar{1}0)$ plane and we rotated the other plane along the $[11\bar{2}3]$ axis with a rotation angle ~127°. This process generated an asymmetric GB with $\Sigma = 5$, and its triclinic cell has following lattice parameters: $a = 7.79$ Å, $b = 20.12$ Å, $c = 46.05$ Å, $\alpha = 137.37°$, $\beta = 50.68°$, and $\gamma = 105.19°$. The total number of atoms in the simulation cell for GB III is 277.

The Vienna Ab-initio Simulations Package (VASP) was used to perform all DFT calculations [73,74]. The Kohn-Sham equations were solved using the projected-augmented wave (PAW) method [75,76], along with standard PAW potentials for the elements Zn and O. The Perdew-Burke-Ernzerhof (PBE) [77] exchange-correlation functional was utilized to perform the DFT optimizations for all three GB structures. The lattice parameters of ZnO GB cells were kept unchanged during the structural relaxation and atomic positions were subjected to relaxation. The Brillouin-zone integrations were sampled on a $\Gamma$-centered 2×2×1 $k$-point grids for all GB structures. The kinetic energy cutoff for plane waves was set to 400 eV and the convergence criterion for electronic self-consistency was adopted to $10^{-4}$ eV. All GB atoms were fully relaxed until the Hellmann-Feynman forces were smaller than 0.02 eV/Å.

**4.5 *Ab initio* molecular dynamic (AIMD) simulations**

Following with the DFT-optimized GB structures, we performed *ab initio* molecular dynamics (AIMD) simulations under the *NVT* ensemble with a Nose-Hoover thermostat to calculate GB diffusivities. The temperature was set at 1573 K (1300 °C), close to the experimental condition. The overall simulation time was set to 2000 fs with a time step of 1 fs. Although the overall simulation time is short (due to the limitation of very large simulation cells), the GB structure can reach equilibrium based on monitoring the potential energy vs. time. The $k$-point grid was 1×1×1 ($\Gamma$ point only). The atoms' trajectories during AIMD simulations were used to calculate the mean squared displacement (MSD) over time (*t*). Finally, the GB diffusivities for Zn and O were obtained by linearly fitting the corresponding MSD *vs. t* curves for both stoichiometric and oxidized (Zn-deficient) GBs for comparison. The average GB diffusivities were calculated based on five different trajectory paths during the AIMD simulations, and the standard deviations were used as the error bars.

**CRediT authorship contribution statement**

**Jian Luo:** Conceptualization; Formal analysis; Funding acquisition; Investigation; Methodology; Project administration; Resources; Supervision; Validation; Visualization; Writing - original draft; Writing - review & editing. **Qizhang Yan:** Data curation; Formal analysis; Investigation; Visualization; Writing - original draft. **Chongze Hu:** Data curation; Formal analysis; Investigation; Visualization; Writing - review & editing.

**Declaration of Competing Interest**

The authors declare that they have no known competing financial interests or personal relationships that could have appeared to influence the work reported in this paper.

**Acknowledgements**

This work is supported by the Materials of Extreme Properties program of the U.S. Air Force Office of Scientific Research (AFOSR) under the grants nos. FA9550-19-1-0327 (for 2019-2022) and FA9550-22-1-0413 (for 2022-2027) and we thank our program manager, Dr. Ali Sayir, for his support and advice. We also thank Dr. Jiuyuan Nie for his assistance and suggestions on setting up the experiments.

**Author contributions**

J.L. conceived the idea and supervised the work. Q.Y. conducted the experiments. C.H. conducted the DFT and AIMD simulations. All authors analyzed the data and contributed to the final manuscript.

**Data Availability**

All research data required to reproduce these findings are available in the manuscript and the Supplementary Data (Supplementary Figures S1-S16).

**Appendix A. Supplementary Data**

Supplementary data to this article can be found online at:

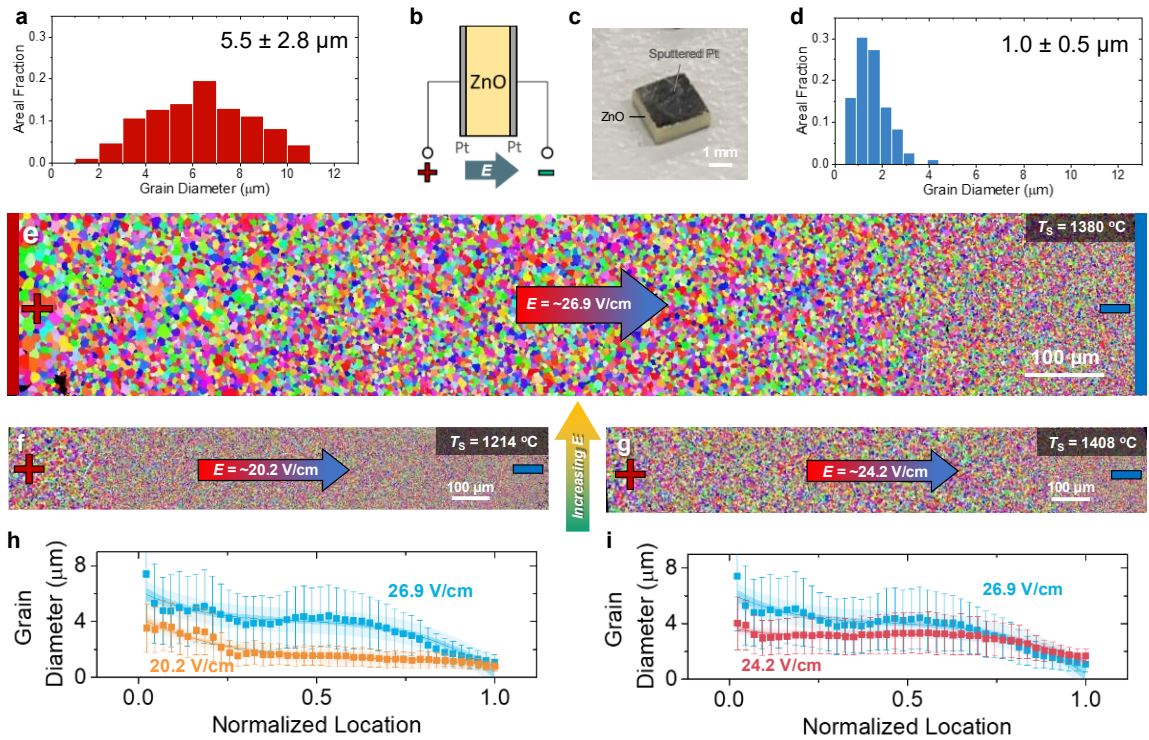

**Figure 1. Creating and controlling graded microstructures with electric fields**. The distribution of grain sizes at regions near (a) the anode and (d) the cathode, respectively. (b) A schematic illustration of the experimental setup. (c) Photo of a dense pristine ZnO specimen with two surfaces sputtered with Pt as electrodes. (e) Electron backscatter diffraction (EBSD) map of a ZnO specimen annealed under a nominal electric field of ~26.9 V/cm at an estimated specimen temperature of ~1380 °C. The graded microstructures can be altered by (f) reducing the nominal electric field to ~20.2 V/cm at a fixed furnace temperature of 600 °C (with a reduced estimated specimen temperature of ~1214 °C due to less Joule heating) or (g) selecting the conditions (different furnace temperatures to compensate different Joule heating) to keep similar estimated specimen temperatures (~1408 °C *vs.* ~1380 °C) with a reduced nominal electric field of ~24.2 V/cm (to show true field effects). Comparisons of the grain size distributions across the specimens under different applied electric fields at (h) the fixed furnace temperature and (i) similar estimated specimen temperatures. The error bars in the plots show the standard deviations of the grain sizes in the corresponding vertical section of the EDSB maps.

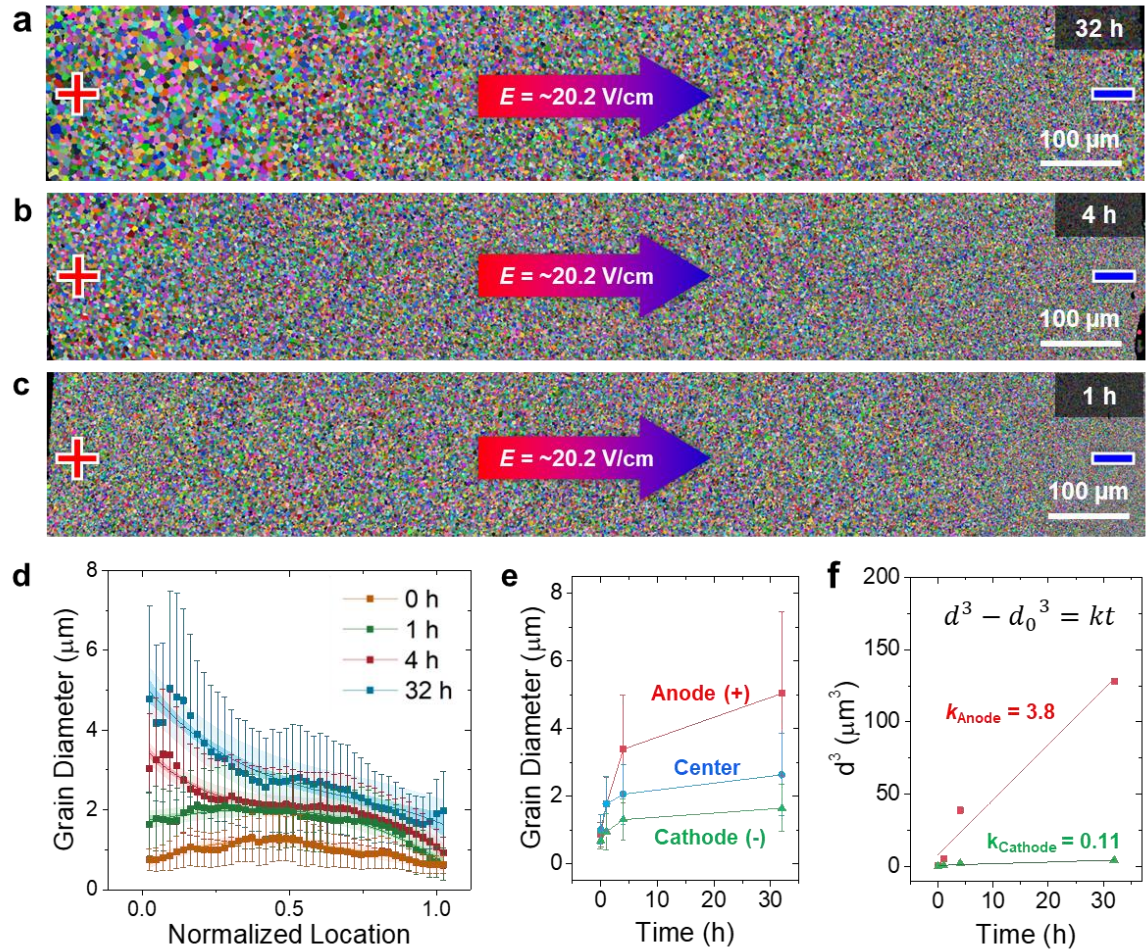

**Figure 2. A kinetics study of the microstructural evolution of ZnO annealed under a constant applied voltage.** EBSD maps of the ZnO specimens annealed for (a) 32 h, (b) 4 h, and (c) 1 h, respectively, under a constant applied voltage of 2.80 V. (d) Grain size distributions of the specimens annealed for different durations under this same applied voltage. The error bars show the standard deviations of the grain sizes of the corresponding vertical sections of the EBSD maps. (e) The averaged grain sizes near the anode, at the center, and near the cathode of the specimens annealed for different time periods. (f) Fitting of grain growth kinetics at the anode *vs.* cathode side, showing a >30× greater grain growth kinetic constant *k* at the anode side than that at the cathode.

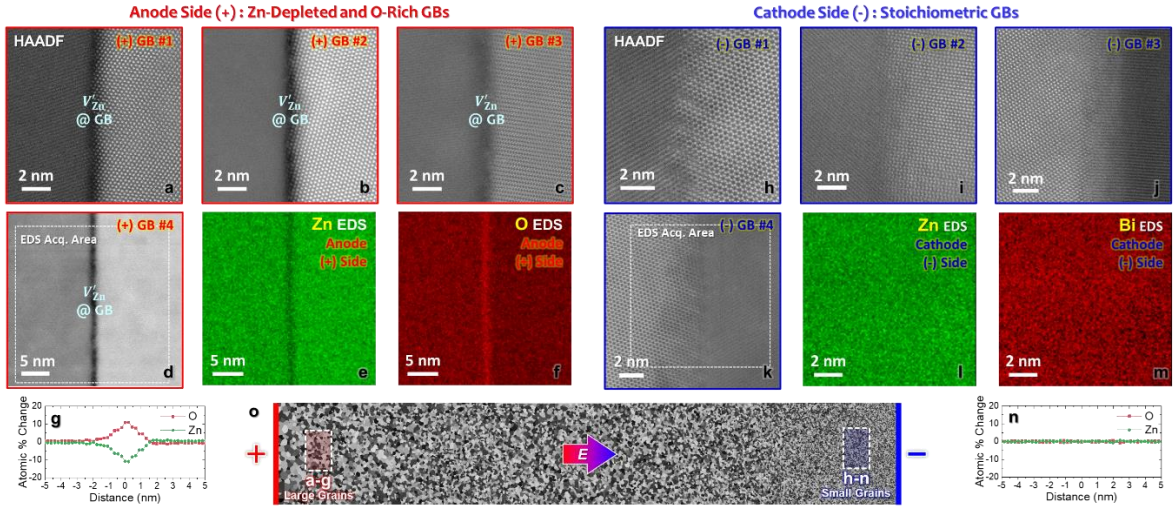

**Figure 3. Aberration-corrected scanning transmission electron microscopy (AC STEM) and X-ray energy dispersive spectroscopy (EDS) characterization of anode- *vs.* cathode-side grain boundaries (GBs).** Oxygen-rich and Zn-deficient GBs in the anode (+) side, as evident by: (a-c) STEM high-angle annular dark-field (HAADF) images (that show characteristic dark contrasts at the GBs due to Zn deficiency), (d) an EDS acquisition area, the corresponding (e) Zn and (f) O maps, and (g) compositional line profiles. Stoichiometric GBs at the cathode (-) side, as evident by: (h-f) STEM HAADF images (that lack characteristic dark contrasts at the GBs due to Zn deficiency), (k) an EDS acquisition area, the corresponding (e) Zn and (f) O maps, and (b) compositional line profiles. (o) An EBSD map showing the approximate locations where the focused ion beam (FIB) specimens were lifted. AC STEM HAADF image and EDS elemental maps of the fifth randomly selected grain boundary (GB #5) on the anode side of the specimen are shown in Supplementary Figure S9, further confirming the generality of the formation of Zn-deficient and O-rich GBs on the anode side.

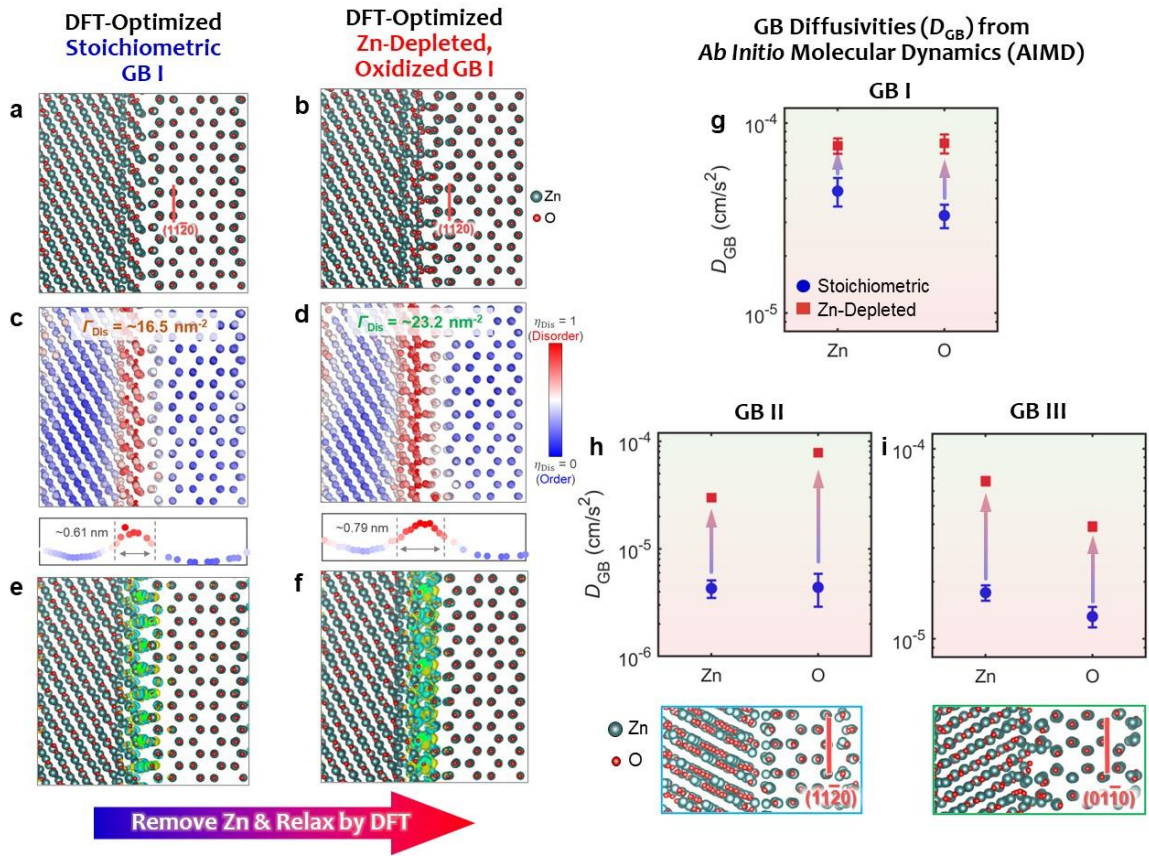

**Figure 4. Density functional theory (DFT)-optimized GB structures and *ab initio* molecular dynamic (AIMD) simulations of GB diffusivities**. DFT-optimized (a) stoichiometric and (b) oxidized (Zn-deficient) GB structures. The calculated disorder parameters for all atoms in the DFT-optimized (c) stoichiometric and (d) oxidized GB structures. The calculated differential charge densities for (e) stoichiometric and (f) oxidized GB structures. For each AIMD simulation, we calculated the average GB diffusivities of Zn and O atoms based on five different trajectory paths for each GB structure. GB diffusivities calculated by AIMD simulations for three independent GBs: (g) GB I shown in Panels 1-f, as well as (h) GB II and (i) GB III. AIMD-simulated mean squared displacement (MSD) *vs.* time profiles and potential energy *vs.* simulation time curves are documented in Suppl. Figs. S11-S16.

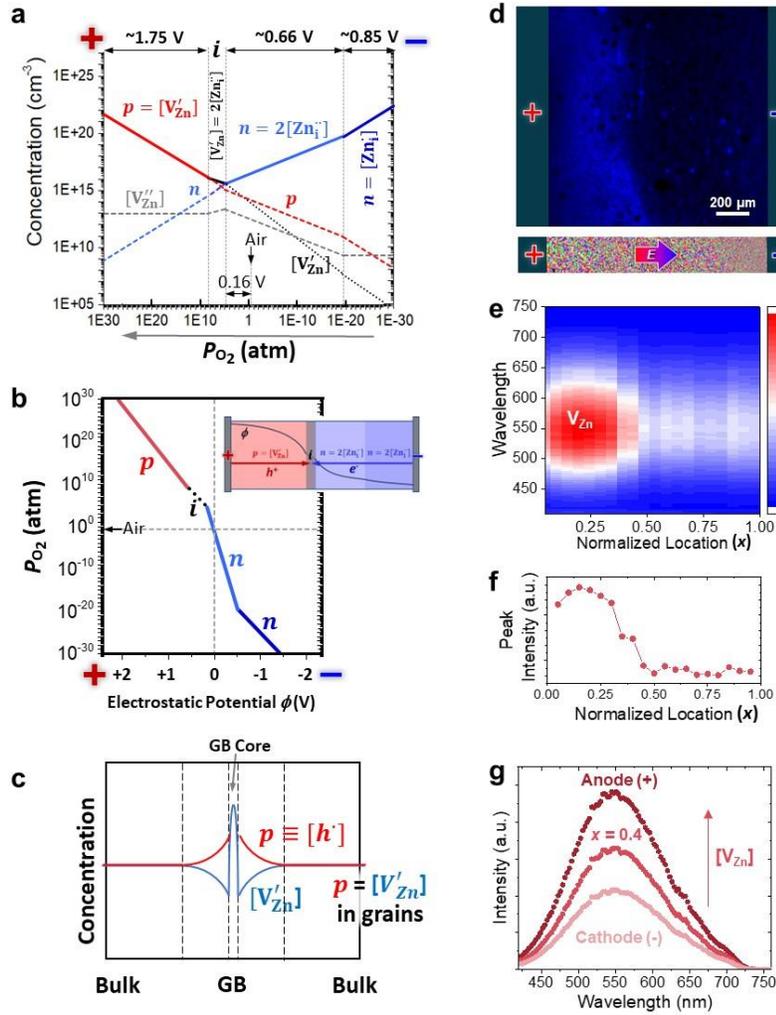

**Figure 5. Defects redistribution in ZnO annealed under electric fields.** (a) Defect concentrations at different oxygen partial pressures in undoped ZnO at 1300 °C following Sukkar and Tuller [36], where only electrons, holes, and Zn vacancies are plotted for figure clarity. The dominant defects are represented by the solid lines and the charge neutrality conditions are given for each region. The electrostatic potential drops were estimated for the *p* and *n* regions and labeled on the top of the diagram. (b) Estimated oxygen partial pressure in the specimen as a function of electrostatic potential. We placed $P_{O2}$ = 0.2 atm at the center of the graph, which may not correspond to the center of the specimen. The actual $P_{O2}$ profile can shift since we do not know the $P_{O2}$ values at the two surfaces of the specimen. The inset shows a schematic illustration of the formation of a *p-i-n* junction in the specimen. However, it is important to note that the actual spatial potential-dependent profile from the anode to the cathode, including the exact location of the *p-i-n* junction, cannot be determined due to the unknown boundary conditions ($P_{O2}$ values) at the two surfaces of the specimen. (c) Schematic illustration of defect concentrations near an anode-side GB. (d) Photoluminescence intensity map at the 550 nm wavelength (representing the concentration of Zn vacancies) of a ZnO specimen annealed under an applied voltage of 2.80 V (over 1.44 mm specimen thickness) for 32 h. The (e) heat map of the 20 integrated spectra across the sample and (f) peak intensity *vs.* normalized position in the specimen (anode: *x* = 0; cathode: *x* = 1), showing the enrichment of Zn vacancies near the anode. (g) Selected photoluminescence spectra at the anode side, near the center at a normalized location of *x* = 0.4 from anode side, and at the cathode side.